\documentclass[
    10pt,
    conference,
    compsocconf
]{IEEEtran}
\pdfoutput=1
\usepackage{xcolor}
\usepackage{graphicx}
\usepackage[utf8]{inputenc}
\usepackage[OT1]{fontenc}
\usepackage[cmex10]{amsmath}
\usepackage{booktabs} 
\usepackage[ruled]{algorithm2e} 
\usepackage{listings}
\usepackage{cleveref}
\usepackage{hyperref}
\crefname{lstlisting}{listing}{listings}
\Crefname{lstlisting}{Listing}{Listings}
\newcommand{\brc}[1]{{\relsize{-1}(#1)}}

\usepackage{fancyhdr}
\fancypagestyle{copyright}{%
    \fancyhf{}
    \cfoot{\footnotesize%
        978-1-5386-2715-0/17/\$31.00~\copyright~2017 IEEE. Personal use of this
        material is permitted. Permission from IEEE must be obtained for all other
        uses, in any current or future media, including reprinting\slash republishing
        this material for advertising or promotional purposes, creating new collective
        works, for resale or redistribution to servers or lists, or reuse of any
        copyrighted component of this work in other works. This material is referenced
        by DOI: \href{https://doi.org/10.1109/BigData.2017.8258495}{10.1109/BigData.2017.8258495}
    }
}

\begin{document}
\title{Event Clustering \& Event Series Characterization on Expected Frequency} 
\author{%
\IEEEauthorblockN{Conrad M Albrecht, Marcus Freitag, Theodore G van Kessel, Siyuan Lu, Hendrik F Hamann}
\IEEEauthorblockA{\\
Physical Analytics\\
TJ Watson Research Center, IBM Research\\
Yorktown Heights, NY, U.S.A.\\
\{cmalbrec,mfreitag,tvk,lus,hendrikh\}@us.ibm.com
}
}

\maketitle
\begin{abstract}
    We present an efficient clustering algorithm applicable to one-dimensional data
    such as e.g.\ a series of timestamps. Given an expected frequency $\Delta T^{-1}$,
    we introduce an $\mathcal{O}(N)$--efficient method of characterizing $N$ events
    represented by an ordered series of timestamps $t_1,t_2,\dots,t_N$. In practice,
    the method proves useful to e.g.\ identify time intervals of \textit{missing}
    data or to locate \textit{isolated events}. Moreover, we define measures to
    quantify a series of events by varying $\Delta T$ to e.g.\ determine the quality
    of an Internet of Things service.
\end{abstract}
\thispagestyle{copyright}

\tableofcontents

\section{Motivation}
The concept of \textit{Internet of Things} \brc{IoT} \cite{itu-t_overview_2012,chiang_fog_2016}
is intimately related to records of certain events, e.g.\ a network
attached device capturing weather information to be broadcasted to other devices
for processing. Given a) the transmitter frequently sends out such information every 
time interval $\Delta T$, and b) the receiving device keeps track of the timestamps
when data was transmitted\slash recorded, the time series $t=\{t_i\}_{i=1\dots N}$
stores information on failure of recording\slash sending\slash transmission\slash
receiving.

If we cluster the one-dimensional data $t$ such that consecutive events are not
more than $\Delta T$ apart, we can infer periods in time where data might be missing.
Upon detection, corresponding action such as retransmission, data interpolation,
etc.\ can be performed. Moreover, the characteristics of \textit{intervals of no
data} \brc{relative frequency, duration, \dots} might help to diagnose the sanity
of the communication network.

Since general purpose, multi-dimensional clustering methods such as Fisher's
discriminant \cite{fisher_use_1936}, \textit{k}--means \cite{lloyd_least_1982}
or more generally \textit{EM} \cite{dempster_maximum_1977} do not exploit the
special property of ordering in one dimension, we aim at a simpler approach that
does not need knowledge of the number of clusters\slash intervals, and it avoids
density estimation such as with \textsc{denclue} \cite{hinneburg_efficient_1998}.

Regarding cluster classification our approach is close to the conceputal notion
\textsc{dbscan} \cite{ester_density-based_1996} introduces: clusters of points and outliers\slash noise points.
However, we exploit the fact that the sequence of timestamps is naturally ordered\footnote{%
    the number of seconds passed since some defined event \brc{for \textit{UNIX epoch time}
    this is Jan 1, 1970 UTC} monotonically increases, thus records of consecutive
    events $1$, $2$, \dots have ordered timestamps $t_1$, $t_2$, \dots}
and thus minimize computational complexity by a factor of $\mathcal{O}(\log N)$.
Of course, if we would have to sort $t$ first, e.g.\ using \textsc{heapsort}
\cite{wegener_worst_1992}, we are back to asymptotic runtime of $\mathcal{O}(N\log N)$.

Our main contribution here is to adapt the concept of \textsc{dbscan} to event
clustering for application in IoT service quality characterization. We present
an algorithm with linear runtime complexity which asymptotically outperforms
native \textsc{dbscan} that operates at an overall average runtime complexity of
$\mathcal{O}(N\log N)$. Of course, while \textsc{dbscan} can be applied to any
number of spatial dimensions our approach is limited to the one-dimensional case.

\section{One-Dimensional Clustering}
\label{sec:1DClusterAlg}

\subsection{Problem Formulation}

Given a set of $N$ ordered timestamps $t=\{t_i\}_{i=1\dots N}$, i.e.
\begin{align}
    i\leq j
    \quad\Rightarrow\quad
    t_i \leq t_j
    \label{eq:OrderCond}
\end{align}
and an \textit{expected} time interval $\Delta T$, provide time intervals $\tau_k$
such that
\begin{align}
    t_i,t_{i+1} \in \tau_k=[\tau_k^-,\tau_k^+]
    \quad\Rightarrow\quad
    \delta t_i=t_{i+1}-t_i \leq \Delta T
    \quad.
    \label{eq:ClusterCond}
\end{align}
Note that $\Delta T$ might be an external parameter to the algorithm that provides
the solution or it is defined by $t$ itself, e.g.\ through $\langle \delta t\rangle=\tfrac{1}{N}\sum_i \delta t_i$.
As we will discuss, the $\delta t_i$ need to be computed and therefore $\langle\delta t\rangle$
is efficiently determined along the lines.

\subsection{Central Idea}
\label{sec:CoreAlg}

In order to fulfill \cref{eq:ClusterCond}, of course, we need to compute at least
$N-1$ time intervals
\begin{align}
    \delta t = \{\delta t_i=t_{i+1}-t_i\}_{i=1\dots N-1}
    \quad.
    \label{eq:Defdt}
\end{align}
Whenever a new time series point $t_{N+1}>t_N$ gets \brc{randomly} added, there
is no a priory way of determining whether $\Delta T$ got exceeded from the existing
$t_{i\leq N}$.

To classify the $t_i$ as interval bounds $\tau^\pm_k$ we note that the binary sequence
\begin{align}
    b = \{b_i=int(\delta t_i>\Delta T)\}_{i=1\dots N-1}
\end{align}
switches from \texttt{1} to \texttt{0} for an \textit{opening} interval bound $\tau^-$,
and from \texttt{0} to \texttt{1} for a \textit{closing} interval bound $\tau^+$, only.
$int(\cdot)$ denotes the function $int(True)\to1$ and $int(False)\to0$.
Hence the quantity
\begin{align}
    B = \{B_i=b_i-b_{i-1}\}_{i=2\dots N-1}
    \quad\text{with}\quad
    B_i\in\{-1,0,1\}
    \label{eq:IntervalBoundClass}
\end{align}
yields the desired association
\begin{align}
    B_i = \pm1
    \quad\Rightarrow\quad
    t_i \in \tau^\pm
    \quad.
\end{align}
Per requirement, a) \cref{eq:OrderCond}, the $t_i$ are ordered, and b) the
binary \brc{discrete} function $b_i$ implies the \textit{alternating property}
\begin{align}
    \forall i<j:~~1=B_i=B_j
    \quad\Rightarrow\quad
    \exists i<l<j:~~B_l=-1
    \quad.
    \label{eq:AltProp}
\end{align}
Thus, linearly scanning through the $t_i$ and their corresponding $B_i$ results in
the two sets
\begin{align}
    \tau^\pm&=\{\tau^\pm_k: k<k' \Rightarrow \tau^\pm_k < \tau^\pm_{k'}\}_{k=1\dots K^\pm\leq N/2-1}
\end{align}
such that we can simply interleave these to obtain the corresponding time intervals
as our solution, \cref{eq:ClusterCond}.

\subsection{Boundary Conditions}
\label{sec:BoundCond}

However, there is a couple of options how to exactly interleave the $\tau^\pm$ which
depend on the boundary condition. More specifically, let us assume the sequence $t_1,t_2,\dots$
starts e.g.\ with intervals that are smaller than $\Delta T$. In this case, $\tau^-_1>\tau^+_1$,
and one needs to manually add a $\tau^-_0$ to construct intervals
\begin{align}
    \tau_k=[\tau^-_{k-1},\tau^+_k]
    \quad.
\end{align}
A corresponding issue might happen at the end of the time series $\{t_i\}$ depending
on whether $K^+=\left\vert\tau^+\right\vert$ is equal or not equal\footnote{
    Note that by virtue of \cref{eq:AltProp} the difference $\left\vert K^+-K^-\right\vert$
    is at most $1$. Actually, it is already obvious from the fact that $\vert B\vert=N-2\neq\vert t\vert=N$
    that one needs to manually add $\tau_k^\pm$ -- imagine the case where each $t_i$
    is a boundary value, but we have two of the $B_i$ missing to classify all $t_i$.
} to $K^-=\left\vert\tau^-\right\vert$.

\begin{lstlisting}[
    float=t,
    caption = {
        Sample implementation of our clustering procedure as pseudo-code.
    },
    label   = {lst:PseudoCodeImpl}
]
algorithm cluster_events is
    input:  #\sf list# t #\sf of ordered timestamps $t$#,
            #\sf float variable# dT #\sf of expected inverse frequency $\Delta T$#
    output: #\sf list# tau #\sf of cluster intervals $\tau$#,
            #\sf list# x #\sf of isolated timestamps $x$#

    define #\sf lists# tauMinus, tauPlus, tau, x
    define #\sf lists# dt, b, B

    N #$\gets$# #\sf length of# t
    b[0] #$\gets$# 1
    b[N] #$\gets$# 1

    for each i in 1,2,#\dots#,N-1 do
        dt[i] #$\gets$# t[i] - t[i-1]
        if dt[i] #$>$# dT then b[i] #$\gets$# 1
        else b[i] #$\gets$# 0

    for each i in 0,1,#\dots#,N-1 do
        B[i] #$\gets$# b[i+1] - b[i]
        if B[i]   = -1  then #\sf append \tt t[i] \sf to \tt tauMinus# else
        if B[i]   = 1   then #\sf append \tt t[i] \sf to \tt tauPlus#   else
        if b[i+1] = 1   then #\sf append \tt t[i] \sf to \tt x#

    for each i in 0,#\dots#,#\sf length of# tauMinus do
        #\sf add interval \tt [tauMinus[i], tauPlus[i]] \sf to \tt tau#

    return tau, x
\end{lstlisting}

In order to prevent manually dealing with all the \brc{four} different boundary
condition scenarios, we might want to \brc{virtually} add the following timestamps
from the outset:
\begin{align}
    t_0 = -\infty
    \quad\text{and}\quad
    t_{N+1}=+\infty
    \quad.
\end{align}
Hence, we obtain $\delta t_0=\delta t_N=+\infty$,
and therefore
\begin{align}
    b_0 = b_N = 1
\end{align}
which yields $N$ $B_i$ that corresponding to $N$ $t_i$ for classification such
that we always have
\begin{align}
    \tau=\left\{\tau_k=[\tau^-_k,\tau^+_k]\right\}_{k=1\dots K}
    \label{eq:ClusterSol}
\end{align}
from
\begin{align}
    \tau^\pm&=\left\{\tau^\pm_k: k<k' \Rightarrow \tau^\pm_k < \tau^\pm_{k'}\right\}_{k=1\dots K}
\end{align}
with $\vert\tau\vert=K\leq N/2$.

\subsection{Isolated Points}
\label{sec:IsolPoints}

Given the solution \cref{eq:ClusterSol}, due to the ordering of the $t_i$, we can
simply form the open intervals
\begin{align}
    \bar{\tau}=\left\{\bar{\tau}_k=\left(\tau^+_k,\tau^-_{k+1}\right)\right\}_{k=1\dots K-1}
\end{align}
that we associate with \textit{time intervals of failure}.  Note, that $[t_1,t_N]=\tau\cup\bar{\tau}$.
However, these intervals do \textbf{not} imply
\begin{align}
    \forall i,k: t_i\notin\bar{\tau}_k
\end{align}
i.e., informally, it is not true that no event happens during the intervals $\bar{\tau}$,
but we certainly have
\begin{align}
    t_i\in\bar{\tau}_k
    \quad\Rightarrow\quad
    \vert t_i - t_{i\pm1} \vert > \Delta T
\end{align}
where we refer to $t_i$ as an \textit{isolated} event. In terms of \textsc{dbscan}
these timestamps form the \textit{noise}, while all $t_i\in\tau^\pm$ are
\textit{border points}.

Isolated events have $b_i=1$ and since they are not interval boundary
points they need to have $B_i=0$. This way we can use $b$ and $B$ to classify
isolated events according to
\begin{align}
    B_i = 0 \wedge b_i = 1
    \quad\Rightarrow\quad
    t_i\in x
\end{align}
where $x$ denotes the set of \textit{isolated} timestamps. Likewise, we can define
\textit{clustered} timestamps as
\begin{align}
    B_i=0 \wedge b_i = 0
    \quad\Rightarrow\quad
    t_i\in \bar{x}\quad.
\end{align}
Since $b$ is binary and \cref{eq:IntervalBoundClass} holds for $B$, all $t_i$ are
uniquely classified, i.e.\ $t=\tau^+\cup\tau^-\cup x\cup\bar{x}$.
It is rather straightforward to convince oneself that there is the association
$\stackrel{(-)}{x}\leftrightarrow\stackrel{(-)}{\tau}$ in the sense that
all $t_i\in x$ are within an interval of $\bar{\tau}$ and $t_i\in\bar{x}$ within
an interval of $\tau$.

\subsection[Implementation \& Comp.\ Complexity]{Implementation \& Computational Complexity}

\Cref{lst:PseudoCodeImpl} provides an example implementation of the method from
\cref{sec:CoreAlg,sec:BoundCond} in pseudo-code for demonstration purposes.
E.g.\ the call of \texttt{cluster\_events(t, dT)}
on\begin{center}\tt\relsize{-.5}
    t=[-20,-18,1,2,2.9,10,11,100,200,202,202,203]
\end{center}
given \texttt{dT} as\begin{center}\relsize{-.5}
    \texttt{-1, 0, 1, 10, 100}, and mean of the elements of \tt t
\end{center}
returns output equivalent to\newline{\tt\relsize{-2}
~[], [-20, -18, 1, 2, 2.9, 10, 11, 100, 200, 202, 202, 203]\newline
~[[202, 202]], [-20, -18, 1, 2, 2.9, 10, 11, 100, 200, 203]\newline
~[(1, 2.9), (10, 11), (202, 203)], [-20, -18, 100, 200]\newline
~[(-20, -18), (1, 11), (200, 203)], [100]\newline
~[(-20, 203)], []\newline
~[(-20, 11), (200, 203)], [100]
}\newline
respectively.

The procedure presented in \cref{sec:CoreAlg,sec:BoundCond,sec:IsolPoints} and \cref{lst:PseudoCodeImpl}
uses $N-1$ algebraic operations for $\delta t$, $N-2$ logical operations for $b$
and again $N$ algebraic operations for $B$ which determines the interval boundary
classification with a total of $3(N-1)$ operations. The final loop in \cref{lst:PseudoCodeImpl}
to interleave the \texttt{tauPlus} and \texttt{tauMinus} lists is just for the
user's convenience.

The naive approach would compute two time intervals for each $t_i$ and perform two
logical operations of those against $\Delta T$ to determine the classification,
hence $4N$ computations. Note that due to the given linear ordering in one-dimensional
space, our algorithm's runtime $3N-3$ is exact. In particular, it is fully deterministic
when the number of timestamps $N$ is fixed.

Moreover, the required memory for our approach is linear in $N$. Only the event
series list \texttt{t} of size $N$ and the lists \texttt{x}, \texttt{tauMinus}, and \texttt{tauPlus} with
a total size of at most $N$ timestamps need to be stored. The lists \texttt{dt},
\texttt{b}, and \texttt{B} can be computed on the fly occupying storage $\mathcal{O}(1)$.

\begin{figure}[t!]
    \includegraphics[width=.5\textwidth]{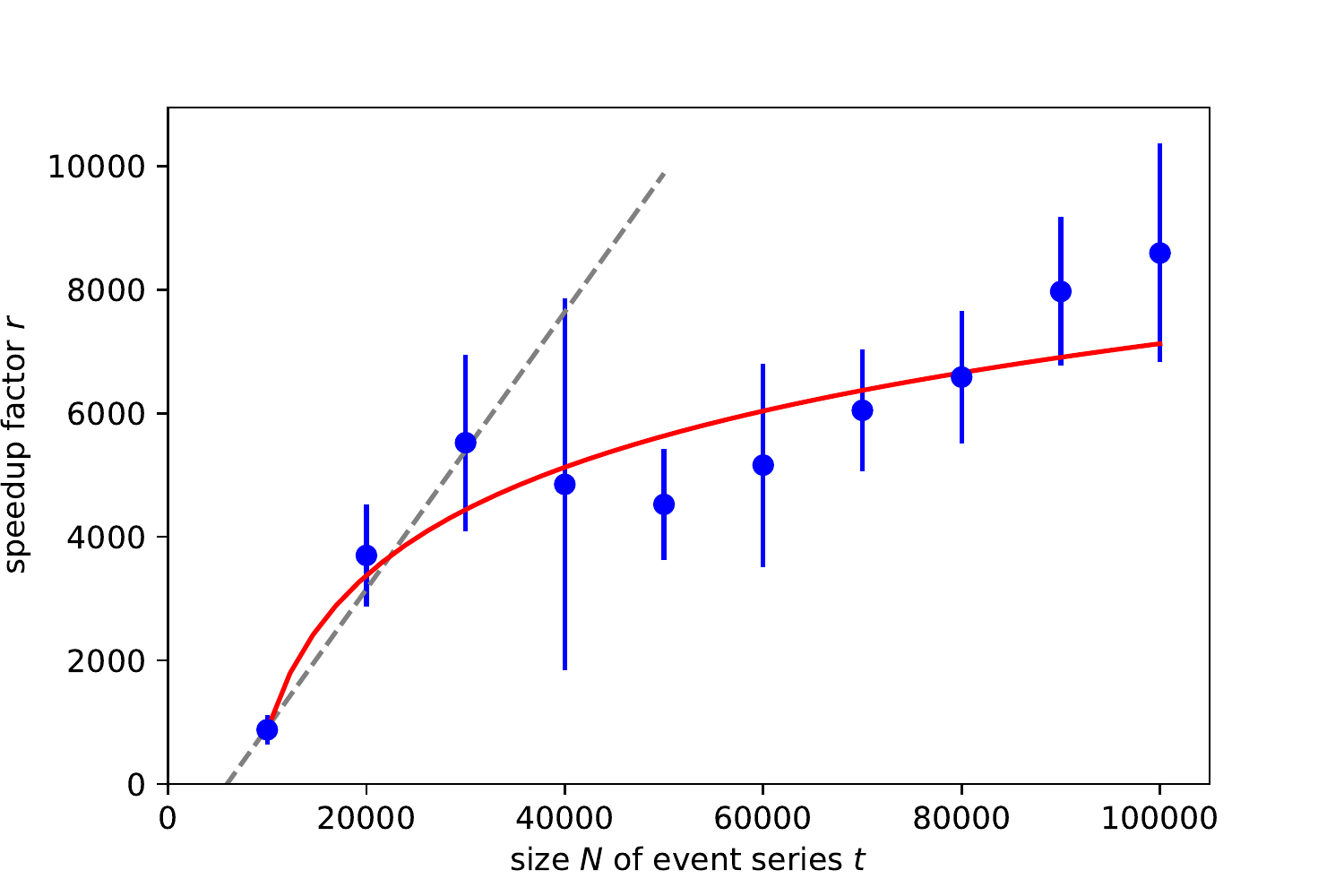}
    \caption{\label{fig:SpeedupCf2DBSCAN}
    Numerical speedup analysis comparing a Python implementation of our algorithm
    using the module \texttt{numpy} \cite{walt_numpy_2011}, v1.13.0, to the vanilla implementation
    of \textsc{dbscan}, \texttt{DBSCAN()}, from the package \texttt{sklearn} \cite{pedregosa_scikit-learn:_2011},
    v0.18.1, module \texttt{cluster} with parameter settings \texttt{min\_samples=2}
    and \texttt{metric='l1'}. We plot the quantity $r$ \brc{cf.\ blue dots} from
    \cref{eq:SpeedUp} versus the number of timestamps $N=\vert t\vert$ to cluster.\newline
    Timing measurements were performed with the standard Python module \texttt{timeit}
    on commodity hardware with sufficient RAM to prevent swapping. Experiments have been
    repeated 40 times to aggregate statistics for error estimation using error propagation
    by first order Taylor expansion.\newline
    The event series was generated by a white noise random distribution on $[0,1)$
    and has been rescaled by $N$. For the experiments, \texttt{dT} was set to two values:
    \texttt{1} and \texttt{1e-4}, corresponding to the parameter \texttt{eps} of
    \texttt{DBSCAN()}.\newline
    While we observed an approximately linear speedup for small $N$ \brc{cf.\ gray,
    dashed fit line $a\cdot r+b$}, an overall logarithmic speedup \brc{cf.\ red, solid fit line
    $c\log[a\cdot r+b]$} is plausible which supports the analytical result:
    $\mathcal{O}(N\log N)/\mathcal{O}(N)=\mathcal{O}(\log N)$.
    }
\end{figure}
To confirm our analytical findings we performed a numerical experiment which is
presented by \cref{fig:SpeedupCf2DBSCAN}. It evaluates the speedup of our algorithm
compared to a vanilla implementation of \textsc{dbscan}. Within the observed error
boundaries, the scaling factor $\mathcal{O}(\log N)$ is plausible for large $N$
wrt.\ the \textit{speedup factor}
\begin{align}
    \label{eq:SpeedUp}
    r(N) = R_\text{\sc dbscan}(N) / R_\text{lin}(N)
\end{align}
with $R_\text{\dots}$ the individual runtime of \textsc{dbscan} and our linear
approach, respectively.

\subsection{Application}

We observe that for a given, fixed event series $t$ with total time interval
$\Delta t=t_N-t_1$, the quantity
\begin{align}
    \label{eq:NoFailureMeasure}
    C^o_t(f)
    = \frac{1}{\Delta t}\sum_k\vert\tau_k\vert
    = \begin{cases}
        1       & \Delta T\leq 0\\
        0\dots1 & 0<\Delta T< \Delta t\\
        0       & \Delta T\geq\Delta t
    \end{cases}
\end{align}
where
\begin{align}
    f = \log\left[\Delta T^{-1}/(\Delta t/\vert t\vert)^{-1}\right]
    = -\log\Delta T\vert t\vert/\Delta t
\end{align}
computes the fraction of time with \textit{no failure in operation}. $\Delta T$
is fixed by the \textit{expected}, logarithmic, and normalized \textit{event frequency}
$f$, i.e.\ $f=0$ represents the scale of frequency where all timestamps are equally
spaced within the time series interval. $f>0$ corresponds to smaller scales, $f<0$
to larger ones.

Scanning $C_t^o$ by varying $f$ provides a characteristics that quantifies the reliability
of e.g. an IoT service. It is rather straightforward to show that $C^o_t$ is monoton
decreasing with $f$ increasing\footnote{
    The larger $\Delta T$, the more the clusters $\tau$ cover the whole time series.
    Due to \cref{eq:ClusterCond} clusters never shrink in size for increasing $\Delta T$,
    they either grow or merge to bigger clusters, letting the overall cover increase.
}.

In case where the time series is generated by a single, periodic data stream,
we get a unit step function $C_t^o(f)=\Theta(-f)$, i.e.\ $1$ for $f<0$ and $0$ for
$f>0$. Nevertheless, similar information could be obtained by simply checking a histogram
$n(\delta t)$, cf.\ \cref{eq:Defdt}, that counts the number of $\delta t_i$ in some
binning interval \brc{number density}. In the case above we would observe a single
peak in $n(\delta t)$. Note, that $C_t^o$ contains similar information to
$\tfrac{1}{\Delta t}\int_0^{\Delta T}n(\delta t)d\delta t$.

\begin{figure}[t!]
    \includegraphics[width=.5\textwidth]{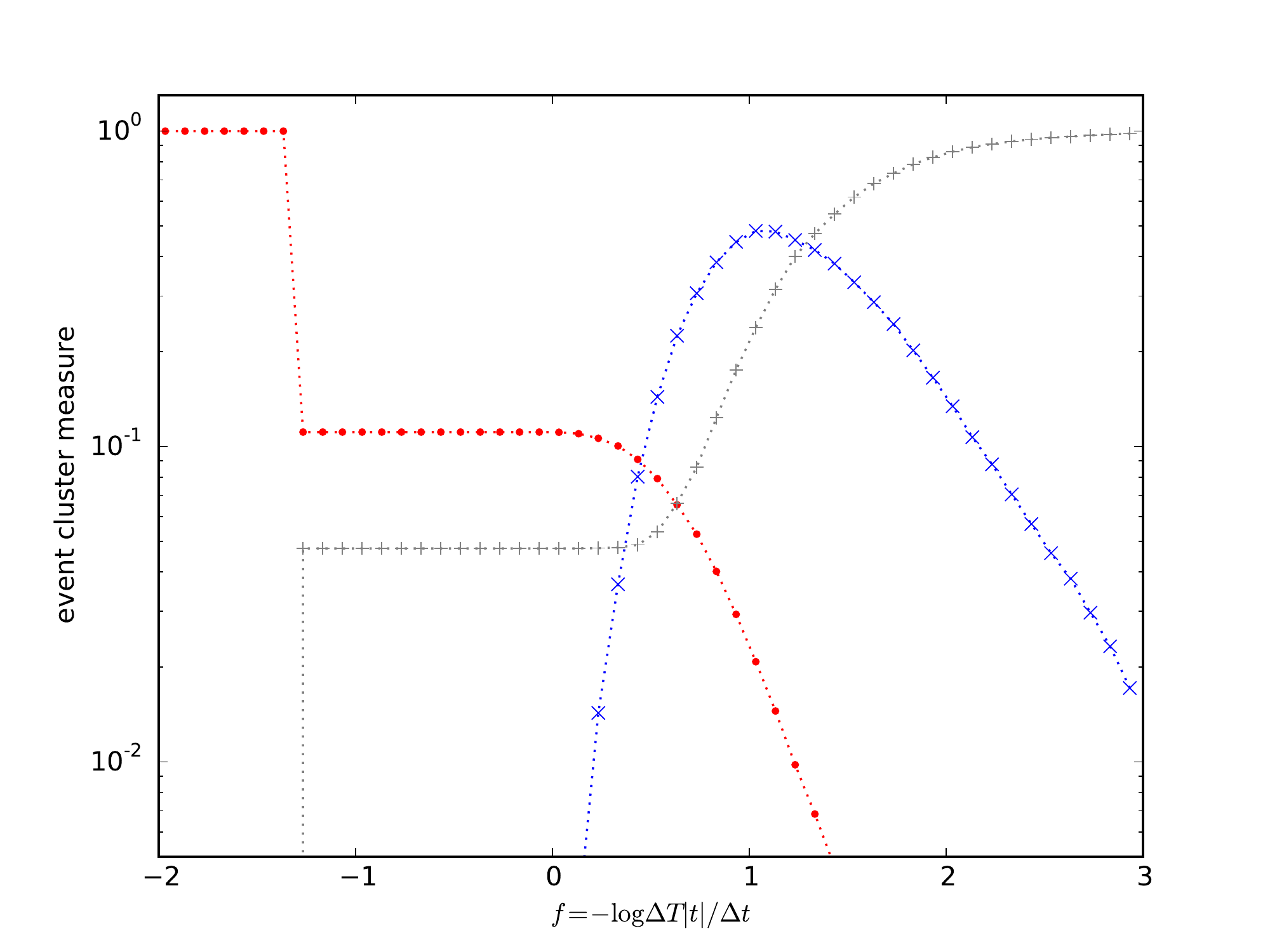}
    \caption{\label{fig:RandTimeSeriesEx}
        Sample plot of IoT service quality measures from event series $t$:
        $C_t^o$ \brc{$\cdot$, red}, $C_t^n$ \brc{$\times$, blue}, and
        $C_t^s$ \brc{$+$, gray} by varying $\Delta T$. The series $t$ consists
        of a burst of random events that covers approx. 10\% of the total time range
        $\delta t$. During the rest of the time the events are periodic at a rate
        of about $1/50\Delta t$.}
\end{figure}

However, our clustering output $(\tau,x)$ provides information that $n(\delta t)$
is blind to, because it does not account for the ordering of the $\delta t_i$.
In particular,
\begin{align}
    \label{eq:ClusterNumberMeasure}
    C^n_t(f)
    = 2\frac{\vert \tau\vert-\delta_{1\vert\tau\vert}}{\vert t\vert}
    = \begin{cases}
        0       & \Delta T< 0\\
        0\dots1 & 0\leq\Delta T< \Delta t\\
        0       & \Delta T\geq\Delta t
    \end{cases}
\end{align}
provides a normed measure of the number of clusters\footnote{
    The Kronecker delta $\delta_{ij}$ is $1$ for $i=j$, $0$ else. It forces
    $\vert\tau\vert\in\{0,1\}$ to result in $C_t^n=0$.
}. While $C^o_t$ just quantifies the total coverage of $t$ by the clusters, $C_t^n$
provides insight whether the coverage is established by a number of patches or a
single\slash a few intervals with data frequency of at least $\Delta T^{-1}$. This
way we might draw conclusions on e.g. the reliability of an IoT service. Ideally
we want $C^n_t\ll1$.

Last but not least, we might consider the number of isolated events
\begin{align}
    \label{eq:IsolatedEventsMeasure}
    C^s_t(\Delta T)
    = \frac{\vert x\vert}{\vert t\vert}
    = \begin{cases}
        0       & \Delta T< 0\\
        0\dots1 & 0\leq\Delta T< \Delta t\\
        1       & \Delta T\geq\Delta t
    \end{cases}
\end{align}
as an additional indicator of reliability, since they are orthogonal to the information
contained in $\tau$. We might classify isolated events as indicator of loose IoT
service quality and thus it should stay close to zero until it quickly increases
to one for some $f>0$.

\Cref{fig:RandTimeSeriesEx} illustrates these applications by plotting $C_t^{o,n,s}$
for an event series $t$ generated from $10^4$ uniformly random samples drawn from
$[0,1]$ joined by $10^3$ equi-distant samples in $[1,10]$. We observe that at $f=0$
there is little variance in $C_t^{o,s}$, indicating that there is no single dominant
event frequency $\nu_0=\vert t\vert/\Delta t$. Moreover, there is a step in $C_t^o$ at
$f\approx-1$ that covers 90\% of its range which refers to a dominant event frequency
one order of magnitude lower than $\nu_0$. Since $C_t^n\ll1$ we conclude this frequency
to be present along major time intervals within $[t_1,t_N]$. Also, $C_t^s$ rapidly
drops. Therefore, the existence of isolated events vanishes at time scales larger
than $\sim\Delta T/10\nu_0$ such that we have a \textit{clean signal}.

In contrast, $C_t^n\sim 1$ for $f\approx1$. Thus, due to the randomness we introduced
in our sample, for high-frequency events, increasing coverage of $[t_1,t_N]$ is achieved
by a number of isolated clusters \brc{random nature of the signal!}. Finally, for
frequencies 3 orders of magnitude larger than $\nu_0$, $C_f^s\approx1$, i.e.\ no
more clustering of events is present.

\begin{figure*}[!t]
    \includegraphics[width=\textwidth]{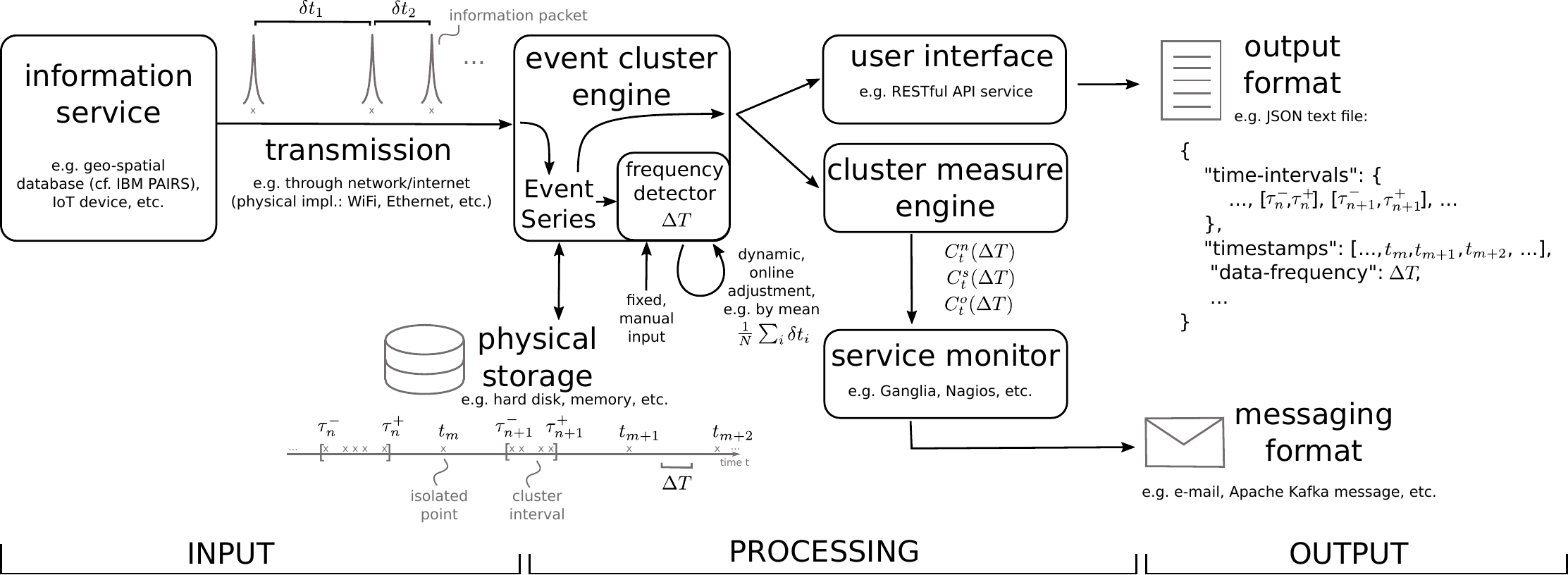}
    \caption{\label{fig:DataProcFlow}
        Sample data and processing flow of an implementation of the IoT quality service
        employing the procedure \cref{lst:PseudoCodeImpl} as well as the measures
        \cref{eq:NoFailureMeasure,eq:ClusterNumberMeasure,eq:IsolatedEventsMeasure}.
        The main text provides details. Note, that here the $t_m$, $t_{m+1}$ and
        $t_{m+2}$ reference isolated timestamps of the set $x$, not to be confused
        with all timestamps $t_i$ of $t$ indexed by $i$, i.e. $x\subseteq t$.
    }
\end{figure*}

\Cref{fig:DataProcFlow} depicts a sample data flow and processing pipeline where
the discussed method can be employed to rate and monitor e.g.\ an IoT device or
the data availability of satellite imagery in the big geo-spatial data platform
\href{https://pairs.res.ibm.com}{\textsl{IBM PAIRS}} \cite{klein_pairs:_2015,lu_ibm_2016}.
Given that this \textit{information service} is expected to send data packages at
frequency $\Delta T^{-1}$, an \textit{event cluster engine} records and stores the
timestamps $t_i$ for further analysis. At the same time a \textit{frequency detector}
might dynamically adjust $\Delta T$, e.g. by computing
the mean of the $\delta t_i$ over a given time window. The event clustering engine
is coupled to a \textit{user interface} that might be interacted with by a \textsl{RESTful API}
\cite{fielding_architectural_2000} served by e.g.\ \textit{Python Flask} \cite{wikipedia_flask_2017}
to trigger the execution of \cref{lst:PseudoCodeImpl} in order to return the sets $\tau$ and $x$.
Once the clustering has been performed, the quantities $C_t^{o,n,s}$ can be computed
and analyzed by a \textit{cluster measure engine} which itself feeds  derived service
quality indicators to a monitoring system such as e.g.\ \textit{Ganglia} \cite{wikipedia_ganglia_2017}
or \textit{Nagios} \cite{wikipedia_nagios_2017}. These might then release alerts
by an appropriate messaging service such as plain e-mail or employing a system such
as \textsl{Apache Kafka} \cite{wikipedia_apache_2017}.

\section{Conclusion}

We discussed and implemented a one-dimensional, one-parameter clustering method
with linear complexity on input and memory usage. It might be the preferred choice
over the more general apporach \textsc{dbscan} takes when clustering ordered timestamps.
Based on the algorithm's output we suggested measures that have useful application
in the domain of IoT to quantify data availability or to indicate the reliability\slash
stability of an IoT device connecting to the network. In particular, the presented
approach is part of the data availability RESTful service of IBM's big geo-spatial
database PAIRS.

The cluster method might be useful for other domains as well. Applications that
have to characterize peaks of measurements can benefit. One of them is the problem
of geo-locating leaks through a network of detector sensors \cite{haupt_demonstration_2005}
such as in the field of industrial pollution detection \cite{albertson_mobile_2016}.

\bibliographystyle{IEEEtran}
\bibliography{PAIRS-project,general-computational}

\end{document}